\input epsf.tex
\documentstyle[iopconf1]{article}
\begin{document}

\title{NEUTRINO TRAPPING AND NEUTRINO MASS BOUNDS}

\author{Brian Woodahl, Michelle Parry, Shu-Ju Tu, and 
Ephraim Fischbach}

\affil{Physics Department, Purdue University, 
West Lafayette, IN  47907 USA}

\beginabstract

It has been shown recently that the exchange of virtual $\nu\bar{\nu}$ pairs
leads to an unphysically large energy-density in neutron stars and white
dwarfs, unless neutrinos have a minimum mass, $m_\nu \stackrel{>}{\sim} 
0.4~ \mbox{eV/c}^2$.
Here we consider the possibility that the presence of trapped low-energy
neutrinos can suppress the exchange of virtual $\nu\bar\nu$ pairs,
thereby avoiding a large energy-density even for massless neutrinos.
We show that a) there can be subvolumes in a neutron star or white dwarf
where neutrino-trapping does not take place, and which can thus
have an unphysically large energy density, and b) even in those volumes
where trapping does occur, the resulting suppression can be too small to
alter the conclusion that neutrinos must have a minimum mass.
\endabstract

It is well known that the exchange of massless neutrino-antineutrino 
$(\nu\bar\nu)$ pairs gives rise to a long-range interaction among
electrons, protons, and neutrinos [1-7].  The 2-body potential between
electrons, $V_{ee}^{(2)}(r)$, is given by [1,4,7]
$$  V_{ee}^{(2)}(r) = \frac{G_F^2(2\sin^2\theta_W + \frac{1}{2})^2}
                           {4\pi^3 r^5} 
                    = \frac{3\times 10^{-82}\mbox{eV}}
                            {(r/1\mbox{m})^5} .  
    \eqno{(1)}
$$
Here $r = |\vec{r}_1 - \vec{r}_2|$ is the separation of the electrons,
$G_F$ is the Fermi constant, and $\theta_W$ is the weak mixing angle.
As can be seen from Eq.(1), the 2-body potential arising from
neutrino-exchange is extremely weak, but its effects may nonetheless
be detectable in equivalence principle experiments [6].

Neutrino-exchange can also lead to many-body interactions, 
and these have been studied by Feynman [2]
and Hartle [3].  Among other observations, Feynman noted that
higher order (in $G_F$) many-body interactions could be important
because they depend on higher powers of the masses of the interacting
objects.  It has been shown more recently [7] that the self-energy of
a compact object such as a white dwarf or neutron star arising from
many-body neutrino-exchange interactions can become unphysically
large, when calculated in the standard model.  In the absence of
some suppression mechanism, one is led to the conclusion
that all neutrinos must have a minimum mass, 
$$  \mbox{m}_\nu \stackrel{>}{\sim} 0.4 ~~\mbox{eV/c}^2 .
   \eqno{(2)}
$$

In order to understand more clearly what any suppression mechanism
would have to achieve, we briefly review the argument leading to the
bound in Eq.(2).  The self-energy of a spherical neutron star arising
from the interaction of 4 particles, for example, must be of order
$(1/R)(G_F/R^2)^4$.  Since the number of 4-particle interactions that
can arise among $N$ particles (where $N={\cal O}(10^{57}$)) is given by the
binomial coefficient
$$  \left( \begin{array}{c} N\\ 4 \end{array} \right) =
       \frac{N!}{4!(N-4)!} \cong \frac{N^4}{4!} ,
    \eqno{(3)}
$$
the 4-body contribution $W^{(4)}$ to the total energy
$W = \sum_k W^{(k)}$ is of order 
$$     W^{(4)} \sim \frac{1}{4!} \frac{1}{R} \left(\frac{G_FN}{R^2}\right)^4 .
       \eqno{(4)}
$$
For a typical neutron star $(G_FN/R^2) = {\cal O}(10^{13})$, and hence it
follows that the analogous contributions from the interactions of
$6,8,..$ particles would rapidly become very large, and eventually
exceed the known mass of the neutron star.  For later purposes it
is helpful to note that the neutrino-exchange energy in a subvolume
of neutron star matter of radius $r_o \cong 1.7\times 10^{-5}$ cm would
be greater than the total mass of the neutron star.

In order to reduce the self-energy $W$ to a physically acceptable
value, there must be a mechanism which suppresses the contributions
from neutrino-exchange.  The possibility explored in Ref.[7] is that
neutrinos have a minimum mass (given in Eq.(2)), whose effect is
to ``saturate" the neutrino-exchange force, in analogy to nuclear
forces.  Smirnov and Vissani (SV) [8] have recently proposed another mechanism:
Very low energy neutrinos trapped in the neutron star or white dwarf
may suppress the exchange of the virtual neutrinos (via the Pauli
principle) which give rise to the neutrino-exchange force. 
Neutrino trapping has been considered by Loeb [9], and more recently
by Kiers and Weiss [10].

In qualitative terms, trapping can occur when the kinetic energy of
neutrinos in a medium is less than their binding energy in the
medium, which is of order $\sqrt{2} G_F\rho \cong 50~ \mbox{eV}$.
(Since the sign of the potential energy for $\nu$ in a medium is
opposite to that for $\bar\nu$ to order $G_F$, $\nu$ can be bound
in the medium while $\bar\nu$ is repelled.)  Under these circumstances
a neutrino can have a total energy which is negative, and hence
cannot escape from the medium to empty space where its total energy
would be positive (see Fig.\ 1).
\begin{figure}
{\hskip98pt \epsfxsize=4cm \epsfbox{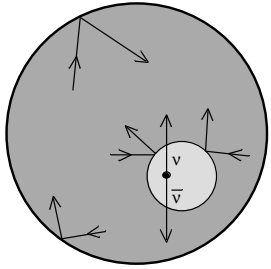}}
\caption{Anti-trapping of neutrinos in a ``bubble".}
\end{figure}
Trapping can be described either in
terms of a variable index of refraction in the medium [9], or as
the quantum-mechanical scattering of neutrinos from a potential [10].
In either picture, however, there is no guarantee that a significant
number of neutrinos will be trapped in every subvolume of the neutron
star, especially one as small as $r_o$.  Consider, for example,
the ``bubble" region shown in Fig. 1, {\it i.e.}, a region of relatively
low density which may form temporarily.
Neutrinos from outside this region cannot enter the ``bubble" for
the same reason that they cannot escape into empty space.
Contrariwise, neutrinos produced inside the ``bubble" {\it can}
escape the ``bubble" region.  The ``bubble" can thus be viewed
as a region of ``anti-trapping", {\it i.e.}, where there are few
trapped neutrinos but where the neutron density is nonetheless large enough
to produce an unphysically large neutrino-exchange energy.

Another argument against the SV mechanism focuses on the time required
for the weak interactions to produce an equilibrium distribution of
trapped neutrinos.  Suppose that the matter distribution in a neutron
star suddenly changes, either due to a ``starquake" or to the accretion
of additional matter.  The time $\tau_1$ required for the many-body potential to
build up in a region of dimension $R$ is of order $R/c$, which
for $R = r_o = 1.7 \times 10^{-5}$ cm is 
$\tau_1 \cong 6 \times 10^{-16}\mbox{s}$.
This estimate follows by noting that in the Schwinger formalism,
the neutron-star ``vacuum" produces a $\nu\bar\nu$ loop which
simultaneously couples to all the neutrons located in a region of
size $R \cong c \tau_1$, via many-body interactions.
By contrast, the time required for the weak interactions to produce
an equilibrium distribution of low-energy neutrinos via
elastic neutrino scattering [11] is of order $\tau_2 = (\sigma_{\nu n}\rho c)^{-1}$,
where $\sigma_{\nu n}$ is the weak $\nu n$ cross section, and $\rho$
is the number density of neutrons. For
$E_\nu = 10$ eV, $\lambda_n \equiv (\sigma_{\nu n}\rho)^{-1} \cong
10^{10}$ km, and $\tau_2 \cong 4 \times 10^{4}$s.  Other estimates
for the $\tau_2$ give similar results.  We conclude from this
argument that the many-body neutrino-exchange potential, and its
associated energy density, comes into being on a very much
shorter time scale than does the trapped neutrino distribution.
Hence a distribution of trapped neutrinos will not always be present to
suppress many-body neutrino exchange.
This discussion is sufficient to demonstrate that one cannot
rely on the presence of trapped neutrinos to suppress the large
neutrino-exchange energy.  In what follows it is shown that even in
regions where trapped neutrinos are present they do not necessarily suppress
the effect calculated in Ref.\ [7] sufficiently to avoid the conclusion
that there is a lower bound on neutrino mass.

We begin by summarizing the argument of Smirnov and Vissani [8].
In the notation of Ref.[7] the vacuum propagator for a massless
neutrino of energy $E$ is
$$  S_F^{(0)} (\vec{r}_{ij},E) = \gamma\cdot \eta
      \left[ \frac{i}{4\pi} \frac{e^{i|E|(r+i\epsilon)}}
                                  {r+i\epsilon} \right]
     \eqno{(5)}
$$

\noindent
where $r = |\vec{r}_{ij}| = |\vec{r}_i-\vec{r}_{j}|$, and
$\gamma\cdot \eta \equiv \vec{\gamma}\cdot \vec{\partial} - \gamma_4 E$.
Eq.(5) leads directly to the 2-body potential $V^{(2)}(r)$ between
neutrons arising from neutrino exchange, which is given by
Eq.(1) with $(2\sin^2\theta_W + 1/2) \rightarrow a_n$, where
$a_n = -1/2$ is neutrino-neutron coupling constant in the standard
model.  In the presence of a background neutrino sea 
at low temperature,
the expression
in square brackets Eq.(5) acquires an additional contribution given
by
$$  n_\nu \frac{\sin|E|(r+i\epsilon)}
               {2\pi r+i\epsilon} .
    \eqno{(6)}
$$
The neutrino distribution function $n_\nu$ is given by

$$  n_\nu = \frac{\theta(E)}{e^{(|E|-\mu)/k_BT} + 1}
        \stackrel{T\rightarrow 0}{\longrightarrow}
         \theta(E) \theta(\mu -|E|),
   \eqno{(7)}
$$

\noindent
where $k_B$ is the Boltzmann constant, 
$T$ is the temperature, 
and $\mu$ is the 
strength of the potential trapping the neutrinos. 
The effect of the additional term proportional to
$n_\nu$ is to replace $V^{(2)}$ by $\tilde{V}^{(2)}$ where [5]
$$   \tilde{V}^{(2)}(r) = \frac{G_F^2 a_n^2}{4\pi^3r^5}
      [\cos (2\mu r) + \mu r \sin(2\mu r)] .
    \eqno{(8)}
$$

\noindent
Smirnov and Vissani then argue that when $\tilde{V}^{(2)}(r)$ is
integrated over a spherical matter distribution such as a neutron
star, the arguments of the cosine and sine functions will
oscillate rapidly, so that $\tilde{U}^{(2)} = \int d^3 r\tilde{V}^{(2)}(r)$
will average to zero.  Since the same considerations would apply to
all the $k$-body potentials $\tilde{V}^{(k)}$
$(r_{12}, r_{13}, r_{14},...)$, the implication is that the total
energy $W = \sum_k W^{(k)}$ would be small, even for massless
neutrinos.

At the heart of the above argument is the Riemann-Lebesque (RL)
theorem [12]:
If $\int_a^b d\theta \psi(\theta)$ exists, and if $\psi(\theta)$
has limited total fluctuation the interval $(a,b)$ then for
$\lambda \rightarrow \infty$
$$  I(\lambda) \equiv \int_a^b d \theta \psi(\theta) \sin (\lambda\theta)
    = {\cal O}(1/\lambda).
   \eqno{(9)}
$$
In the limit $\lambda \rightarrow \infty$, $I(\lambda) \rightarrow 0$
which establishes the theorem in its oft-used form.
[We note that the RL theorem ensures that $I(\lambda)$
tends to zero {\it at least as fast as} $1/\lambda$.  It may turn out
that coefficient of the $1/\lambda$ term vanishes, so that $I(\lambda)$
tends to zero even faster than $1/\lambda$, {\it e.g.} as $1/\lambda^2$.]
In the present context $\lambda = \mu R$, where $R = 10 $ km is the
radius of the neutron star, and $\mu = \sqrt{2} G_F \rho = 50$ eV.
It follows that $1/\lambda = 1/\mu R\cong 10^{-12}$ is indeed small,
but is not zero.  Moreover, $1/\lambda$ multiplies a factor
$(G_F\rho R)$ which is itself very large.  Hence one cannot establish
without a more careful analysis whether the $1/\mu R$ suppression is
sufficient to reduce $W$ to a physically acceptable value without
having to introduce a neutrino mass.

To anticipate the results described below, we note from Eq.(8) that
the term containing $\sin(2\mu r)$ has a coefficient which is proportional
to $\mu$.  Since the integration over $r$ introduces a factor
$1/\mu^n$ $(n \geq 1)$ as demanded by the RL theorem, the net
suppression of this term will be of order $1/\mu^{n-1}$.  Hence if
$n=1$ there is no suppression at all of the term proportional to
$\sin(2\mu r)$.  As we demonstrate below, the leading 2-body
contribution $W^{(2)}$ to $W$ does in fact correspond to $n=1$,
so that there is no suppression at all of $W^{(2)}$ from trapped neutrinos.

Following the discussion of Ref.[7], the average interaction energy 
$\tilde{U}^{(2)}$ of a single pair of neutrons having a uniform probability
distribution in a spherical neutron star of radius $R$ is given by
$$  \tilde{U}^{(2)} = \int_{r_c}^{2R} dr {\cal P}(r) \tilde{V}^{(2)} (r).
    \eqno{(10)}
$$
Here $r_c$ is the neutron hard-core radius, and ${\cal P}(r)$ is the probability
density for finding two points a distance $r$ apart in a sphere of
radius $R$ [7],

$$  {\cal P}(r) = \frac{3r^2}{R^3} - \frac{9}{4} \frac{r^3}{R^4}
        + \frac{3}{16} \frac{r^5}{R^6} .
    \eqno{(11)}
$$
Combining Eqs.(8) and (11) we find

$$  \everymath={\displaystyle}
   \begin{array}{rcl}
      \tilde{U}^{(2)} &=& \frac{3G_F^2a_n^2}{8\pi^3R^3}
        \biggr[ \frac{3}{8R^2} \cos(4\mu R) + \frac{1}{r_c^2} \cos (2\mu r_c) 
           - \frac{3}{2r_cR} \cos(2\mu r_c) \\
         && + \frac{r_c}{16R^3} \cos (2\mu r_c) + \frac{3}{32\mu R^3}
            \sin(4\mu R)\\
        && \\
        && - \frac{3}{32\mu R^3} \sin(2\mu r_c) + \frac{3\mu}{2R}
              \mbox{Si} (4\mu R) - \frac{3\mu}{2R} \mbox{Si}
              (2\mu r_c) \biggr] .
    \end{array}
        \eqno{(12)}
$$
Here Si(z) is the sine integral function defined by
$$ \mbox{Si(z)} = \int_o^z dt \frac{\sin t}{t} .
     \eqno{(13)}
$$
For $\mu r_c \cong 1\times 10^{-7} << 1$ we have
$$  \tilde{U}^{(2)} \cong \frac{3}{8\pi^3} \frac{G_F^2a_n^2}{r_c^2R^3},
       \eqno{(14)}
$$
which is the same result found in Ref.[7] for the case of no neutrino
background $(\mu = 0).$
The 2-body case demonstrates that
even when trapped neutrinos are present their effects
may be too small to significantly modify the conclusions of
Ref. [7].

The results for the many-body contributions $W^{(k)}$ $(k \geq 4)$
can be analyzed in a similar fashion.  In the absence of any effects
due to trapped neutrinos, the 4-body potential $V^{(4)}$ is given
by the expression in Eq.(3.50) of Ref.[7].
For present purposes it is sufficient to study the contribution
from the zero-derivative term $V_o^{(4)}$ in $V^{(4)}$ given by

$$ \everymath={\displaystyle}
  \begin{array}{rcl}
  V_o^{(4)} &=& \biggr(\frac{G_Fa_n}{\sqrt{2}}\biggr)^4
          \frac{3!}{\pi^5} \frac{1}{P_4S_4^5} ,\\
      P_4 &=& r_{12} r_{23} r_{34} r_{41}; ~~~~~~~~~
          S_4 = r_{12} + r_{23} + r_{34} + r_{41} .
  \end{array}
    \eqno{(15)}
$$

The effect of $n_\nu$ in (6) is to replace $V_o^{(4)}$ by
$$  \everymath={\displaystyle}
   \begin{array}{rcl}
      \tilde{V}_o^{(4)} &=& \frac{(G_Fa_n)^4}{2\pi^5P_4S_4}
        \biggr\{ \cos(\mu S_4)\left(\frac{3}{S_4^4}
              - \frac{3\mu^2}{2S_4^2} + \frac{\mu^4}{8}\right)\\
          && \\
         && + \frac{\mu \sin(\mu S_4)}{S_4}
           \left(\frac{3}{S_4^2} - \frac{\mu^2}{2}\right) \biggr\}.\\
    && \\
\end{array}
     \eqno{(16)}
$$
As in the 2-body case, the coefficients of the oscillatory terms
$\cos(\mu S_4)$ and $\sin(\mu S_4)$ contain explicit powers of $\mu$
which offset powers of $1/\mu$ that will arise when
$\tilde{V}_o^{(4)}$ is integrated over a spherical volume.
This integration can be carried out in analogy to the 2-body case by
writing
$$ \tilde{U}_o^{(4)} = \int dr_{12}dr_{23} dr_{34} dr_{41}
          {\cal P}_4(r_{12}, r_{23}, r_{34}, r_{41})\tilde{V}_o^{(4)}
               (r_{12}, r_{23}, r_{34}, r_{41})
    \eqno{(17)}
$$
where ${\cal P}_4(...)$ is the 4-body analog of ${\cal P}(r)$ in Eq.(11).
Even though the functional form of ${\cal P}_4(r_{12},r_{23},r_{34},r_{41})$
is not known, we can infer from the preceding discussion of the 2-body
case that the four integrations in (17) will in general introduce a
suppression factor of $(1/\mu R)^4$, via the RL theorem.
However, this factor will be canceled by the explicit factor of
$\mu^4$ which appears as a coefficient of $\cos(\mu S_4)$ in Eq.(16).
Although it is possible that a term with a  larger suppression
factor, say $(1/\mu R)^5$, might be the dominant contribution,
there is no reason to expect this for a general matter distribution. 
Moreover, even if a suppression from
trapped neutrinos does arise, one cannot ensure that it would be
sufficient to offset the large energy density obtained in Ref. [7].

Many-body effects may be important in other neutrino processes,
as shown in Fig. 2.
\begin{figure}
{\hskip0in \epsfxsize=4.4in \epsfbox{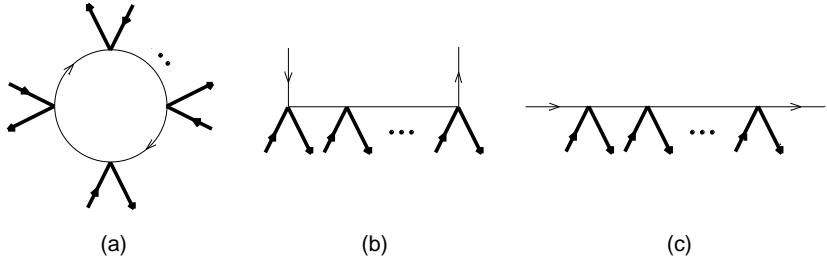}}
\caption{(a) Self-energy of neutrons arising from neutrino
exchange.  Lighter lines denote neutrinos, and heavier lines
are neutrons.  (b) Many-body contributions to $\nu\bar\nu$
production in a medium.  (c) Many-body contributions to neutrino
scattering and the MSW effect.}
\end{figure}
Fig. 2(a) exhibits the $k$-body contribution
to the self-energy of a neutron star which was analyzed
in Ref.[7].  By cutting the neutrino line one obtains Fig. 2(b)
which represents the many-body contribution to $\nu\bar\nu$ production,
and Fig. 2(c) which is the many-body contribution to elastic neutrino
scattering or to the MSW effect [13].  If neutrinos are massless
then all of these diagrams share the same enhancement factor arising from
$k$-body combinatorics, and hence these may be the dominant contributions
to the corresponding processes.  If this were the case, then the
trapping mechanism itself would have to be revisited since for $k$ even
the many-body contributions could have the property of repelling
both $\nu$ and $\bar\nu$.

To summarize, although low-energy neutrinos can be trapped
under some conditions in a neutron star [9,10], one cannot guarantee
that trapped neutrinos will be present in every subvolume of a
neutron star or white dwarf, during every stage in the evolution of
these objects.  In an inhomogeneous medium subvolumes of lower density
(``bubbles") can be regions where ``anti-trapping" occurs:
$\nu\bar\nu$ pairs produced in this region can escape, whereas
neutrinos outside the region cannot enter.  Since such regions
would contain few trapped neutrinos, the mechanism for suppressing
neutrino-exchange forces proposed by Smirnov and Vissani [8] would not
apply.  More importantly, even in regions where neutrinos as trapped,
their effect in suppressing neutrino-exchange forces may be too small
to avoid the necessity for introducing a minimum neutrino mass.

One of the authors (EF) wishes to thank Ken Kiers, 
Abraham Loeb, Jim Pantaleone,
and Alexei Smirnov
for helpful conversations.  We also are indebted to Nancy Schnepp
for help in manuscript preparation.  This work is supported in
part by the U.S. Department of Energy under contract DE-AC02-76ER01428.

\end{document}